\documentclass{iau} 

\usepackage{graphicx}
\usepackage{subfigure}

\newcommand\msun{{M_{\odot}}}
\def\lta{\mathrel{\hbox{\rlap{\hbox{\lower4pt\hbox{$\sim$}}}\hbox{$<$}}}}
\def\gta{\mathrel{\hbox{\rlap{\hbox{\lower4pt\hbox{$\sim$}}}\hbox{$>$}}}}

\title[The self-regulated AGN feedback loop]
{The self-regulated AGN feedback loop: \\ the role of chaotic cold accretion}

\author[M. Gaspari]
{M. Gaspari$^{1}$}

\affiliation{$^1$Department of Astrophysical Sciences, Princeton University, Princeton, NJ 08544, USA
 \\ email: {\tt mgaspari@astro.princeton.edu}; \textit{Einstein} \& \textit{Spitzer} Fellow}

\pubyear{2015}
\volume{319}  
\jname{Galaxies at High Redshift and Their Evolution over Cosmic Time}
\editors{S. Kaviraj, H. Ferguson, eds.}

\begin{document}

\maketitle

\begin{abstract}
Supermassive black hole accretion and feedback play central role in the evolution of galaxies, groups, and clusters.
I review how AGN feedback is tightly coupled with the formation of multiphase gas and the newly probed chaotic cold accretion (CCA). In a turbulent and heated atmosphere, cold clouds and kpc-scale filaments condense out of the plasma via thermal instability and rain toward the black hole. In the nucleus, the recurrent chaotic collisions between the cold clouds, filaments, and central torus promote angular momentum cancellation or mixing, boosting the accretion rate up to 100 times the Bondi rate.
The rapid variability triggers powerful AGN outflows, which quench the cooling flow and star formation without destroying the cool core. The AGN heating stifles the formation of multiphase gas and accretion, the feedback subsides and the hot halo is allowed to cool again, restarting a new cycle. Ultimately, CCA creates a symbiotic link between the black hole and the whole host via a tight self-regulated feedback which preserves the gaseous halo in global thermal equilibrium throughout cosmic time.

\end{abstract}

\vspace{-0.65cm}
\section{Introduction}
The self-regulation mechanism of active galactic nuclei (AGN), i.e., how to link feedback and accretion 
on to supermassive black holes (SMBHs), is matter of intense debate. 
Albeit AGN feedback is supposed to address several astrophysical problems, including
the cooling flow/mass sink problem,
the soft X-ray decline,
the quenching of star formation, and 
the steepening of the scaling relations,
it remains striking that the tiny SMBH can affect
and be affected by the 10-dex larger host (galaxy, group or cluster of galaxies).
This work reviews recent advancements in addressing
the SMBH problem, on both the  
{\it feeding} and {\it feedback} front.
In the feedback problem, we are concerned with the amount of BH energy released into the atmosphere.
Where is the energy deposited? Is the injection driven by the kinetic, thermal, or radiative mode?
Which are the key feedback imprints?
In the feeding problem, we are concerned with the cold or hot mode of accretion on to the BH.
How is the 10 kpc scale linked to the sub-pc scale? Can we go beyond
restrictive Bondi and thin disc formulas? 
In particular, we now know that
the nonlinear interplay of key physics, as turbulence, cooling, and heating,
drives a new accretion mechanism known as \textit{chaotic cold accretion} (CCA). 
Remarkably, once entered in the tight self-regulated loop both feeding and feedback 
play the role of cause and effect: feeding via cold cloud condensation triggers kinetic feedback, 
which can both heat the hot halo and promote thermal instability (TI), further inducing filamentary condensation
and accretion. 

\vspace{-0.2cm}
\section{The feeding problem}
To tackle the fully nonlinear feeding problem, it is necessary to employ hydrodynamic 3D simulations
with very high resolution.
\cite[Gaspari et al.~(2013a, 2015b)]{Gaspari:2013_cca}
systematically studied the impact of turbulence, cooling, heating, and rotation in a common massive galaxy
($M_\ast\simeq3.4\times10^{11}\;\msun$) centrally dominating a galaxy group as NGC 5044 ($M_{\rm tot}\simeq4\times10^{13}\;\msun$). The accretor is a SMBH with $M_\bullet\simeq3\times10^9\;\msun$
and pseudo-relativistic potential.
The adaptive mesh refinement capabilities of FLASH code permit to zoom in the very center, 
resolving the 50 kpc scale down to 0.8 pc -- in a few runs down to 20 gravitational radii ($10^6$ range).
One key physics worth to highlight is turbulence.
The hot atmospheres are continuously shaped by chaotic motions, in analogy to Earth weather.
While waiting for \textit{Astro-H} and \textit{Athena},
we can assess the level of turbulence from relative gas density perturbations, which are linearly related
to the Mach number, $\delta\rho/\rho\approx {\rm Ma_{1D}}$ (\cite[Gaspari et al. 2013c, 2014b]{Gaspari:2013b}). For instance, Coma cluster shows ${\rm Ma}_{\rm 3D}\approx0.45$. AGN feedback, mergers, galaxy motions, and supernovae, all contribute to drive turbulence with at least subsonic velocities, $\sigma_v>100$ km\,s$^{-1}$ (e.g., Sanders \& Fabian 2013).

\vspace{-0.1cm}
\subsection{Hot mode: stifled accretion} \label{s:hot}
The feeding modes can be divided into hot (adiabatic) and cold (radiative) mode.
During adiabatic accretion, in halos dominated either by turbulence (turbulent Taylor number ${\rm Ta_t}\equiv v_{\rm rot}/\sigma_{\rm v}<1$) or rotation (${\rm Ta_t}>1$),
central vorticity, either chaotic or coherent, stifles the accretion rate via a thick centrifugal barrier. Compared with the spherically symmetric \cite[Bondi]{Bondi:1952} rate ($\dot M_{\rm Bondi}\propto M_\bullet^2\,K^{-3/2}$, with $K$ the gas entropy; 1952) -- often blindly applied in literature --
the feeding is suppressed by a factor 3\,-\,4. For supersonic turbulence, the suppression exceeds 1 dex. 
Key point is that in hot, pressure-supported accretion, the accretion rates are very low, e.g.,
$\dot M_\bullet\lta0.01\,\msun\,{\rm yr}^{-1}$ for a massive galaxy.

\vspace{-0.1cm}
\subsection{Cold mode -- CCA: boosted accretion} \label{s:cold}
Hot halos cool through radiative emission.
They shine in X-ray, though the condensed gas can be observed in UV, optical, infrared, and radio (Sec.~\ref{s:obs}).
With weak or no turbulence and rotation, a cooling halo develops a pure cooling flow, i.e., a central monolithic 
condensation with $\dot M_{\rm \bullet}= \dot M_{\rm cool}$.
In a rotating atmosphere, the cooling gas condenses quickly on to the equatorial plane and forms a cold thin disc.
The disc rotational support inhibits feeding, leading to 
$\dot M_\bullet \approx 10^{-2}\,\dot M_{\rm cool}$,
but still $5\times$ higher than the hot mode.

\textit{Chandra} and \textit{XMM} observations have revealed that hot halos are not only perturbed and cooling,
they are also affected by AGN heating (\cite[McNamara \& Nulsen 2012]{McNamara:2012}). 
The volumetric heating rate balances the average cooling rate,
$\mathcal{H}\sim \langle\mathcal{L}\rangle$,
although the AGN injection is anisotropic (e.g., Fig.~9 in \cite[Gaspari et al. 2012a]{Gaspari:2012a}).
Atmospheres shaped by turbulence, cooling, and heating, induce a phenomenon
called {\it chaotic cold accretion} (CCA). The process is familiar to our everyday weather experience.
Cold clouds and filaments condense out of the hot phase and rain down.
The condensation region corresponds to the radius where
$t_{\rm cool}/t_{\rm grav}\lta10$, a few kpc for massive galaxies, up to 10s kpc in cluster cores.
Such threshold is not a line in the sand, showing considerable scatter
(Fig.~10 in \cite[Gaspari et al. 2012a]{Gaspari:2012a}).
As they precipitate, 
the extended filaments, clouds, and torus experience chaotic inelastic collisions, increasing in the inner 100s pc region.
While condensation and turbulence broaden the angular momentum distribution of the cold gas,
collisions promote angular momentum cancellation or mixing. This leads to the rapid accretion (with Myr variability) 
of low-angular momentum cold gas, 
boosting the accretion rate up to $\dot M_\bullet\sim100\,\dot M_{\rm Bondi}\sim \dot M_{\rm cool}$.
As turbulence decays and turbulent Taylor number ${\rm Ta_t}>1$, turbulent broadening, TI, and collisions become weaker,
hence the accretion rate diminishes, approaching the cold disc evolution.
As the disc is consumed, the SMBH only accretes in hot mode with 
highly suppressed accretion rates, $\dot M_\bullet \ll \dot M_{\rm Bondi}$ (Fig.~\ref{fig:1}).

\vspace{-0.1cm}
\subsection{Probing CCA} \label{s:obs}
There are several observational ways to test CCA. In X-ray, the presence or not of CCA can be tested through the plasma radial profiles. CCA model predicts inner flat X-ray temperature profiles due to the recurrent multiphase condensation.
On the other hand, hot accretion models predict cuspy $T_{\rm x}$ profiles due to adiabatic compression. Flat $T_{\rm x}$ profiles and inner multiphase structures have been discovered in NGC 3115 (\cite{Wong:2014}) and M87 (\cite{Russell:2015}); other examples are NGC~4261 and NGC 4472.
Not all galaxies resides in CCA mode; the hot mode is typically associated with more quiescent systems, such as NGC 4649 and NGC 1332 (\cite[Humphrey et al. 2009]{Humphrey:2009}).

\begin{figure*}
        \subfigure{\includegraphics[width=\textwidth]{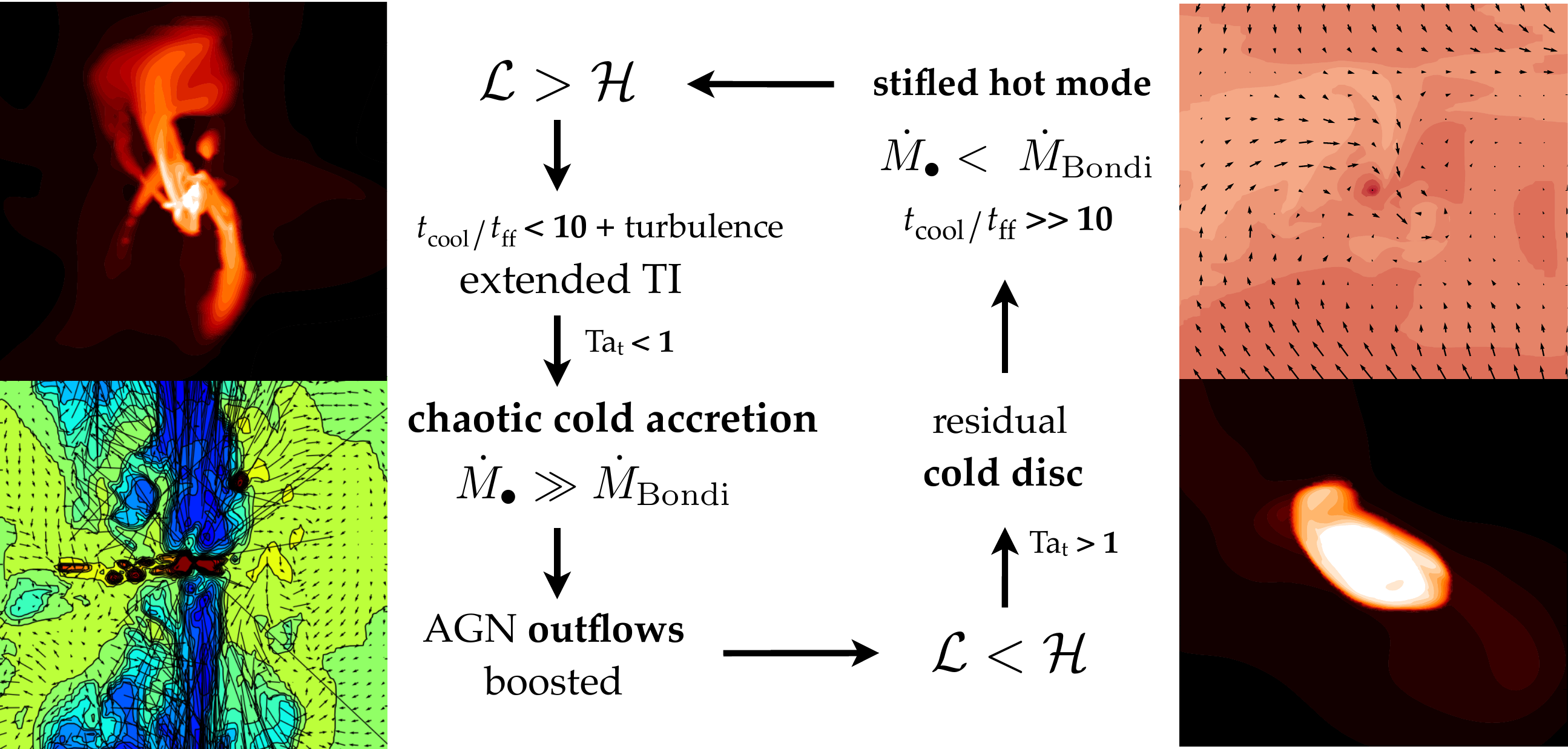}}
        \caption{The self-regulated AGN feedback loop
        (adapted from Gaspari et al.~2013b, 2015b).}
        \label{fig:1}
        \vspace{-0.1cm}
\end{figure*}

A key tracer of residual cold gas and CCA is H$\alpha$ emission.
In many clusters, groups, and galaxies (e.g., Perseus, A1795, NGC 5831, NGC 4636), 
we see extended H$\alpha$ filaments co-spatial with soft X-ray, UV, and molecular emission.
The dominant ionization mechanism is poorly known, including (but not only) gas collisional ionization and photoionization from the stars, AGN, or plasma. 
In the CCA runs, the synthetic H$\alpha$ maps (Fig.~\ref{fig:1}, top left) consistently 
show extended filamentary structures up to 4 kpc.
As ${\rm Ta_t} > 1$, the maps show only a kpc-size cold disc (Fig.~\ref{fig:1}, bottom right).
Remarkable examples of filamentary CCA and disc mode are NGC 5044 and NGC 6868, respectively (\cite{Werner:2014}). 
Further positive observational tests of CCA, e.g., related to star formation and entropy profiles, 
have been presented by \cite[Voit et al.]{Voit:2015} (2015 and references within).

\vspace{-0.35cm}
\section{The feedback problem}
CCA dynamics can be depicted as `raining on to black holes'.
The raining cold clouds and filament can represent the broad-line and narrow-line regions
in AGN observations, 
or high-velocity clouds in Galactic data.
We see the recurrent formation of a central highly clumpy and turbulent torus, 
which is crucial for AGN unification studies.
The chaotic evolution may explain the rapid AGN luminosity variability,
as well as the deflection of jets and changes in the BH spin.
CCA establishes a natural symbiosis between the tiny SMBH and the whole galaxy:
the BH mass results to be linearly related to the galaxy cold gas mass and later to the stars,
$M_{\rm \bullet}\propto M_{\rm cold} \propto M_{\rm \ast}$ (i.e., the Magorrian relation).

The fast communication time
between the SMBH and host, together with boosted accretion, puts CCA as main candidate to drive efficient
and tightly self-regulated feedback.
\cite[Gaspari et al. (2011a,b, 2012a,b, 2014a)]{Gaspari:2011a} 
probed different AGN feedback self-regulation and injection models
in galaxy clusters, groups, and isolated galaxies. These have been tested with large-scale simulations 
(100 pc$\,$-$\,$2 Mpc) for a long-term evolution ($>7$ Gyr), including dark matter and stellar potentials, stellar heating
and mass loss, and gas astrophysics (radiative cooling and AGN heating). The best consistent model resulted in linking the accretion rate to the central cooling -- and not Bondi -- rate, $\dot M_\bullet \sim \dot M_{\rm cool}$, and injecting
the feedback energy in mechanical form via massive subrelativistic outflows.
Self-regulation means $P_{\rm out} \equiv 0.5\,\dot M_{\rm out}\,v^2_{\rm out}=\epsilon\,\dot M_\bullet\,c^2$,
where $\epsilon$ is a mechanical efficiency of the order $10^{-4}$\,-\,$10^{-3}$.
As CCA is triggered (Sec.~\ref{s:cold}; 
$\mathcal{L} > \mathcal{H}$), 
the cold clouds feed the SMBH, boosting accretion and
hence the injected outflow power. The massive outflows gently thermalize the kinetic energy in the core (unlike thermal/Sedov injection), re-heating the core ($\mathcal{L} < \mathcal{H}$), thus quenching the cold gas fueling. 
The gaseous halo is multiphase (hot plasma and warm/cold phase), settling in global thermal equilibrium.
The thermalization of the AGN outflows 
occurs via a combination of weak shocks, inflated 10s kpc bubbles, and turbulence. 
The AGN outflows further induce anisotropic uplift of metals and cold gas. The previous processes are reviewed
in Gaspari et al.~(2013b).

AGN feedback must avoid {\it both} overcooling and overheating, since
consistency with observations implies to quench the pure cooling rates by 20 fold, while preserving the cool-core structure ($\nabla T_{\rm x}(r)>0$). 
More specifically, 
AGN feedback should quench the differential X-ray luminosity below 2 keV.
Inside-out AGN heating can suppress the soft X-ray spectrum by 2 dex compared with the pure cooling flow, 
as outflows deposit more heat in the inner cooler phase (\cite[Gaspari 2015a]{Gaspari:2015_xspec}). 
Turbulence, which becomes transonic in the cooler phase, increases the scatter and further quenches $dL_{\rm x}/dT$.
Overheating is often overlooked. 
Too strong AGN feedback (a thermal/quasar blast) creates a break below 1 keV in the $L_x(r<R_{500})$\,-\,$T_x(r<R_{500})$ relation and morphs every cool core into 
a perennial non-cool-core system (central $t_{\rm cool}>10\, t_{\rm Hubble}$).
This is not observed in X-ray data, as
cool cores populate the universe (e.g.,~\cite{Hudson:2010}).
Anderson et al. (2015) stacked the X-ray emission of 250000 central galaxies, finding no evidence of a break in the scaling relations. The mechanical AGN feedback tightly self-regulated via CCA can consistently preserve cool cores 
for several Gyr, while mildly steepening the $L_x$\,-\,$T_x$  power-law relation.
Figure \ref{fig:1} summarizes such self-regulated AGN feedback loop.

\vspace{-0.14cm}
\acknowledgements
\vspace{-0.13cm}
M.G. is supported by NASA through Einstein Postdoctoral Fellowship (PF-160137) issued by the CXC, which is operated by SAO for and on behalf of NASA (NAS8-03060).



\end{document}